\documentstyle[preprint,tighten,aps,psfig]{revtex}
\begin{document}
\title{Chiral symmetry restoration and the
linear form of baryonic Regge trajectories}
\author{ L. Ya. Glozman}
\address{  Institute for Theoretical
Physics, University of Graz, Universit\"atsplatz 5, A-8010
Graz, Austria\footnote{e-mail: leonid.glozman@kfunigraz.ac.at}}
\maketitle

\begin{abstract} 
It has recently been suggested that the observed 
structure of parity doublets, seen in the spectrum of highly
excited baryons, may be due to the effective restoration
of chiral symmetry for these states. This chiral symmetry
restoration high in the spectrum is  consistent with the
concept of quark-hadron duality. Moreover, if 
 QCD dynamics implies the linear Regge trajectories for highly
 excited baryons, then the MacDowell symmetry requires
the parity doubling to appear, which shows that linear Regge 
trajectories and chiral
 symmetry restoration are well compatible. On the contrary, 
 in the low-energy part of baryon spectrum the parity doublets 
 are absent because of spontaneous chiral
 symmetry breaking. Then the MacDowell symmetry implies that
 there should be no linear parallel Regge trajectories.
 Experimental data shows that this is indeed the case.
\end{abstract}

\bigskip
\bigskip

It is known since 20 years that the low-energy
properties of  nucleon, such as its mass, are
intimately related to the phenomenon of spontaneous
breaking of chiral symmetry  (SBCS) \cite{IOFFE}. At the same
time the physics of the highly excited baryons (hadrons)
is intuitively attributed to the confining property of
QCD. The most significant phenomenological support
to the latter is represented by  the linear-like behaviour of
hadron Regge trajectories. While, in the case of mesons,
the Regge trajectories are indeed approximately parallel
straight lines , this is
by far not true for baryons.
In Fig. 1 (where we have shown only a part of positive and negative
parity Regge trajectories) there are clearly no 
linear parallel trajectories
in the low-energy part of the baryon spectrum (compare trajectories for, 
e.g. the positive parity
$N(939), 1/2^+,$ and negative parity $N(1535),1/2^-$). 
This fact can naturally be
interpreted as that low in the spectrum it is the physics
that is not directly related to confining property of QCD
is crucially important.  Indeed, the whole low-lying
spectrum can be reproduced in the context
of the  chiral constituent quark
model \cite{GR}, where apart from a phenomenological
confining potential the residual interaction between
constituent quarks is mediated by chiral (Goldstone boson)
fields. Both the constituent quarks and chiral fields
can be considered as effective degrees of freedom that
are induced by SBCS.\\

Any constituent quark model in light
baryons that rely on a phenomenological confinement potential 
predicts a lot
of states in the region $M \sim 2$ GeV, that are not observed (the
so-called missing states). The striking
feature of experimental baryon spectrum, however, 
is the systematical appearance of the 
parity doublets in this region. The systematical
parity doubling cannot be explained in  potential constituent
quark model \cite{DOUBLETS} and probably indicates the onset of a new
physical regime. If this phenomenon is experimentally confirmed,
then the most natural explanation would be that a 
parity doubling should be a manifestation of some fundamental
symmetry of the underlying  QCD theory. Recently, it has been
suggested that the observed parity doublets could be explained
as due to the restoration of
almost perfect $SU(2)_L \times SU(2)_R$ global chiral
symmetry of  QCD Lagrangian in the $u,d$ sector \cite{DOUBLETS}.\\

The chiral symmetry is known to be spontaneously broken by the  QCD vacuum,
and, as a consequence, there are no parity doublets  in
the low-energy spectrum. Indeed, in the Nambu-Goldstone mode of chiral symmetry 
 the constituent quarks
are adequate degrees of freedom and the quark model works well
(see, for example, a recent calculation of covariant formfactors 
of the nucleon \cite{FF}).
However, if the effects of SBCS become less
important at increasing energy  in the baryon spectrum ({\it i.e.} chiral
symmetry is effectively restored), the constituent quarks
become inadequate. In this case the physical states should fall
into representations of the chiral group $SU(2)_L \times SU(2)_R$,
that have been called chiral-parity multiplets \cite{COHEN}. Depending
on the specific representation the multiplets can be either parity
doublets in $N$ and $\Delta$ spectra that are not connected to
each other, or  parity doublets that belong to the same
parity-chiral multiplet and hence are degenerate in mass (existing data
probably support the latter case).\\

It has also been suggested in ref. \cite{COHEN} 
that the concept of quark-hadron duality naturally
implies the chiral symmetry restoration.
The phenomenon of quark-hadron
duality \cite{DUALITY} is well established  in many processes, e.g.
in $e^+e^- \rightarrow hadrons$, where we have a direct experimental access
to creation of the quark-antiquark pair by the electromagnetic current.
According to this concept, the spectral density $\rho(s)$
should be dual to the polarization operator calculated
at the free quark loop level (up to perturbative corrections)
at  asymptotically 
large $s$. For the process $e^+e^- \rightarrow hadrons$  the "asymptotic
regime" sets up at $s \sim 2-3$ GeV$^2$.
  In the case of baryons, unfortunately,
there are no experimentally accessible currents that can create three
quarks at some space-time point and connect them to baryons.
Nevertheless, one can construct such currents theoretically \cite{IOFFE},
and these currents are widely  used in QCD sum rules or lattice 
calculations to extract
properties of low-lying baryons directly from QCD. The quark-hadron
duality, applied to the present case, would mean that in the asymptotically
high part of the baryon spectrum the baryon spectral density should be
dual to the one which is calculable in perturbation theory;
hence  the chiral symmetry should be manifest in the spectral
density, because there is no chiral symmetry breaking in
perturbation theory.\\

The pure perturbative contribution is dual to a continuus spectrum
({\it i.e.} very dense  spectrum of overlapping resonances),
so that there should be no
isolated resonances.  Clearly,
as one goes up in the spectrum the nonperturbative contributions
to the spectral density should smoothly decrease, approaching
eventually  regime where a free quark loop is dominant. If the 
nonperturbative effects related
to SBCS die out earlier than those ones responsible for confinement,
then the high-energy part of baryon spectrum should show
isolated baryon resonances that do not feel effects of SBCS
and consequently fall into parity-chiral multiplets.

 So the question arises whether it is possible or not possible
to design nonperturbative effects in such a way that the
spectral density at large $s$ is still nontrivial (i.e. it feels
some nonperturbative contributions) and, at the same time, becomes
chirally symmetric. The answer is probably  positive. 
The aim of this note is to show
that  QCD nonperturbative dynamics that is responsible 
for linear Regge trajectories  in the high-energy
baryon spectrum together with
the assumption of analyticity in relativistic quantum theory
(that is behind Regge-pole phylosophy), is indeed
compatible with effective chiral symmetry restoration.\\  

Let us define the current (interpolating field) $\eta(x)$ that
creates three quarks in a color singlet state at the space-time
point $x$. Then, the two-point correlator
$\langle  0 |T\left ( \eta(x), \bar \eta(0) \right)| 0 \rangle$
describes the propagation of three quarks from the point 0, where they
are created from the vacuum, to the point $x$, where they
annihilate into the vacuum. On their way from 0 to $x$, these three
quarks interact with each other and with the vacuum fields.
This correlation function contains complete information about
all baryons that couple to the current $\eta$.\\ 

To be specific, let us
consider the example of the Ioffe current 
$\eta_N = \epsilon_{abc}\left ( {u^a}^TC\gamma_\mu u^b \right )
\gamma_\mu \gamma_5 d^c$. {\it i.e.} one of the currents that
couples to isodoublet $J=1/2^+$ and $J=1/2^-$ baryons.
Here $a,b,c$ are the colour indices for $u$ and $d$ quarks.
 This current 
couples to isodoublet $J=1/2^+$ and $J=1/2^-$ baryons.
With respect
to   $SU(2)_L \times SU(2)_R$  transformation properties, it
belongs to the $(1/2,0) \oplus (0,1/2)$ representation.  The correlation 
function can be rewritten
in the closure representation as a sum over all possible
baryonic states that can be created by the given current.
Hence, in the zero-width approximation the correlator 
is given by

\begin{equation}
\Pi (q)= i \int d^4x e^{iqx} \langle  0 |T\left ( \eta(x), \bar \eta(0) \right)
| 0 \rangle 
 =\sum_{n_+,n_-}\left[ 
\lambda_{n_+}^2\frac{\gamma_\mu q^\mu+m_{n_+}}{q^2-m_{n_+}^2 
+ i\varepsilon}
+\lambda_{n_-}^2\frac{\gamma_\mu q^\mu-m_{n_-}}{q^2-m_{n_-}^2 + 
i\varepsilon}\right],
\label{corr}
\end{equation}

\noindent
where $m_{n_+}$ and $m_{n_-}$ are masses of positive and negative
parity states and the constants $\lambda_{n_+}$ and $\lambda_{n_-}$
parametrize the strengths with which the given baryonic states
couple to the current:

\begin{equation}
<0|\eta|n_+(q)> = \lambda_{n_+} u(q),
\label{+}
\end{equation}

\begin{equation}
<0|\eta|n_-(q)> = \lambda_{n_-} \gamma_5 u(q),
\label{-}
\end{equation} 

\noindent
where $u(q)$ is a generic Dirac spinor for a baryon with momentum $q$.\\

The correlator (\ref{corr}) contains the chiral even and  odd terms

\begin{equation}
\Pi(q) = \Pi^{even}(q^2)q_\mu \gamma^\mu + \Pi^{odd}(q^2),
\label{EVODD}
\end{equation}

\noindent
which behave differently under chiral transformations;
while the former  conserves chirality, the latter one
violates it. Indeed, under a axial rotation 
$\exp(\imath \pi \tau^3 \gamma_5/2)$ one obtains

\begin{equation}
 \langle  0 |T\left ( \eta(x), \bar \eta(0) \right)
| 0 \rangle = -\gamma_5
\langle  0 |T\left ( \eta(x), \bar \eta(0) \right)
| 0 \rangle \gamma_5.
\label{ROT}
\end{equation}

\noindent
This property implies that in the chiral symmetric phase
the chiral odd term must vanish and
there must be a one-to-one correspondence between the
positive and negative parity baryonic states that are degenerate
in mass and equally coupled to the current:

\begin{equation}
m_{n_+} = m_{n_-},
\label{D1}
\end{equation}

\begin{equation}
\lambda_{n_+} = \lambda_{n_-}.
\label{D2}
\end{equation}

Converse is also valid, the conditions (\ref{D1}) and (\ref{D2})
 imply that the
chiral odd term is zero in the correlation function, which means
that the correlation function is chirally invariant. Actually, this 
is a simple and well known manifestation
of the old statement that if the $SU(2)_L \times SU(2)_R$
symmetry is realized in the Wigner mode, the hadrons should fall
into multiplets of the chiral group that contain degenerate
states of positive and negative parity \cite{PAGELS}.\\

We have considered for simplicity the current that belongs
to the $(1/2,0) \oplus (0,1/2)$ representation. However, one can
similarly construct currents that belong to other representations
of the chiral group \cite{JI}. The current that belongs to the
$(1/2,1) \oplus (1,1/2)$ representation couples at the same
time to both positive and negative parity states with the same
spin in $N$ and $\Delta$ spectra. In the Wigner mode of
chiral symmetry, the chiral odd term in the correlator must vanish, which 
is equivalent to the statement 

\begin{equation}
m_{n_+} = m_{n_-} = m_{\Delta_{n_+}} = m_{\Delta_{n_-}},
\label{D}
\end{equation}

\begin{equation}
\lambda_{n_+} = \lambda_{n_-} = \lambda_{\Delta_{n_+}}
 = \lambda_{\Delta_{n_-}}.
\label{DD}
\end{equation}
\noindent
If the conjecture of Ref. \cite{COHEN}
is correct, then the highly-lying states in $N$ and $\Delta$
spectra should be strongly coupled to the latter current.\\

In the real world, chiral symmetry is spontaneously broken.
 As a consequence, the spectrum does not consist
of systematical parity doublets. For example, it is the 
nonvanishing quark condensate in the vacuum state that
leads to the nonzero value of the chiral odd term in
the correlator and hence
is responsible for the splitting between $N(939),1/2^+$, and
$N(1535),1/2^-$, states (for a recent study in the context of QCD
sum rules see \cite{OKA} and for lattice calculations
see \cite{SASAKI}).
 \\

However, even if the chiral symmetry is spontaneously broken
in the vacuum ( hence, in the low-lying states),
one might expect that it becomes effectively restored 
at higher energies.
Indeed, at large space-like momenta $q^2 < 0$, where the operator
product expansion (OPE) \cite{OPE} is applicable, the correlator is dominated
by free quark loop  plus perturbative 
corrections. The
nonperturbative contributions from the quark condensates,
that are the only contributions to the chiral odd term,
are suppressed by powers of $q^2$. For instance, using
the same Ioffe current like in expressions (\ref{corr})-(\ref{EVODD}), 
the OPE up to dimension dim=6 operators is \cite{IOFFE,RUB}

\begin{equation}
 \Pi^{odd}(q^2) =
 -\frac{1}{4\pi^2} \langle \bar q q \rangle q^2 \ln (-q^2),
 \label{OPEO}
 \end{equation}
 
\begin{equation}
 \Pi^{even}(q^2) = 
 \frac{\ln(-q^2)}{32\pi^2}\left(\frac{q^4}{2\pi^2}  
 + \langle\frac{\alpha_s}{\pi} G^a_{\mu \nu}
G^a_{\mu \nu} \rangle \right) + 
\frac{2}{3q^2}\langle \bar q q \rangle^2 .
 \label{OPEE}
 \end{equation}

\noindent
Hence, upon analytical continuation from the deep
space-like region (where the language of quarks and gluons is
applicable) to the time-like region, $s=q^2 >0$ (where the current creates
baryons), in the high-energy spectral density (that is
imaginary part of the correlator in the time-like region) 
the chiral odd term becomes small
with respect to the chiral even one, because it is suppressed by the same
powers of $s$ in the time-like region as  of $q^2$ in the
space-like one. This implies
that the splitting between parity partners of the high-energy spectrum
will become small with respect to the mass of baryons.\\

High in the spectrum the chiral even spectral density
is 
driven by perturbative
contributions as well as by contributions from gluonic condensate
(the last term in (\ref{OPEE}) does not contribute to the spectral
density).
The gluonic condensate parametrizes soft nonperturbative gluonic
effects in QCD and does not break chiral symmetry. 
If these nonperturbative contributions imply the linear Regge
trajectories  in the  high-energy spectrum, 
then 
they are indeed compatible with the chiral symmetry restoration,
as it will be shown in the following.\\

The position $\alpha(t)$ of the corresponding Regge
pole in  the complex angular momentum plane is identified
with the physical state once the trajectory $\alpha(t)$
crosses the real values $J=1/2, 3/2, ...$ and, at this point,
$t$ is identified with the mass of baryon (for a review of the
Regge theory see  \cite{COLLINS}). A very general requirement
of analyticity implies that fermionic (but not bosonic!) Regge
trajectories of positive and negative parity must satisfy the 
MacDowell symmetry \cite{MACD},

\begin{equation}
\alpha^+(\sqrt{t}) = \alpha^-(-\sqrt{t}), ~t>0.
\label{MD}
\end{equation}

\noindent
Hence, at positive $t$ this equation implies that
the most general functional form
for positive- and negative-parity baryonic Regge trajectories
is given by 

\begin{equation}
\alpha^+(t) = \alpha_0 +  \alpha_1\sqrt{t} + \alpha_2 t + ...,
\label{RP}
\end{equation}

\begin{equation}
\alpha^-(t) = \alpha_0 -  \alpha_1\sqrt{t} + \alpha_2 t - ....
\label{RPM}
\end{equation}

\noindent
Specific QCD dynamics, that is responsible for the given
functional form of Regge trajectories, is hidden in the
coefficients $\alpha_0, \alpha_1, ...$, that in principle
should be calculable in terms of QCD degrees of freedom.
If  QCD nonperturbative dynamics leads to the linear form
 of baryonic Regge trajectories
 in the high-energy part of baryon spectrum, 
 {\it i. e.} $\alpha(t) = \alpha_0 + \alpha' t$,
then at large positive $t$ $\alpha_1 = \alpha_3 = ...=0$ in (\ref{RP}) and
 (\ref{RPM}), {\it i.e.} the MacDowell symmetry requires
that here the positive and negative parity Regge
trajectories must coinside, 

\begin{equation}
\alpha^+(t) = \alpha^-(t).
\label{R}
\end{equation}

\noindent
Hence, it follows 
that baryons will occur in degenerate doublets of opposite parity.
Just the same is implied by the effective chiral
symmetry restoration. One then concludes
that chiral symmetry restoration is consistent with that QCD nonperturbative
dynamics that is responsible for the linear behaviour of Regge
trajectories.\\

On the contrary, in the low-energy spectrum
SBCS 
tells that there must be no degenerate parity doublets,
which is consistent with data. This together with the MacDowell 
symmetry, requires that at small positive $t$
at least one of the coefficients $\alpha_1, \alpha_3, ...$
in (\ref{RP}) and (\ref{RPM}) is not vanishing, which makes
 Regge trajectories curved. Hence the chiral symmetry breaking
effects of QCD together with requirements of analyticity imply 
that there must be no linear parallel
Regge trajectories low in the baryon spectrum. That this indeed the
case is well seen from Fig. 1. \\

In summary, we have shown that the proposed effective restoration
of chiral symmetry in the upper part of the baryon spectrum,
that explains the systematic appearance of parity doublets,
is consistent with the linear form for baryonic Regge trajectories.
On the other hand, the absence of parity doublets in the low-lying
part of the spectrum, that is attributed to spontaneous breaking
of chiral symmetry, implies that there should be no linear
parallel Regge trajectories. The experimental data support it.

{\center{\bf Figure captions}}

\medskip
\noindent
Fig.1 

\medskip
\noindent
Some of the Regge trajectories in the nucleon
spectrum. The solid lines represent the positive parity
Regge trajectories, while the dashed ones - the negative parity
trajectories.

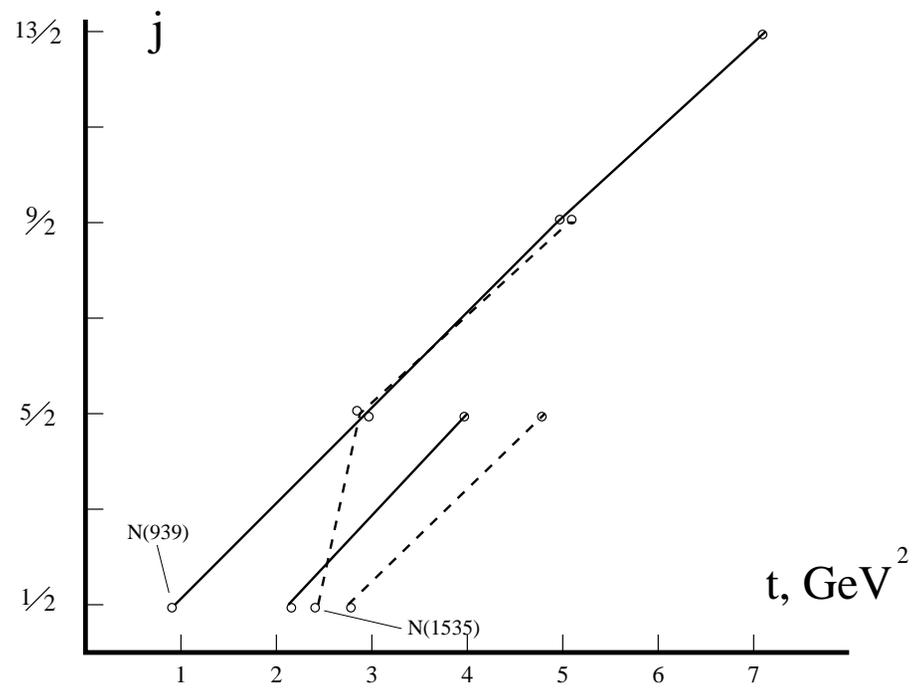
\end{document}